\documentclass[aps,preprint,prd,nofootinbib]{revtex4}

\usepackage{graphicx}
\usepackage{psfrag}
\usepackage{epsfig}

\begin{document}

\title{GENXICC: A Generator for Hadronic Production of the
Double Heavy Baryons $\Xi_{cc}$, $\Xi_{bc}$ and $\Xi_{bb}$}
\author{Chao-Hsi Chang$^{1}$ \footnote{email:
zhangzx@itp.ac.cn}, Jian-Xiong Wang$^{2}$ and Xing-Gang
Wu$^{1,3}$\footnote{email: wuxg@itp.ac.cn}}
\address{$^1$Institute of Theoretical Physics, Chinese Academy of Sciences,
P.O.Box 2735, Beijing 100080, P.R.China.\\
$^2$Institute of High Energy Physics, P.O.Box 918(4), Beijing
100049, P.R.China\\
$^3$Department of Physics, Chongqing University, Chongqing 400044,
P.R. China}

\begin{abstract}
We write down a generator program for the hadronic production of the
double-heavy baryons $\Xi_{cc}$, $\Xi_{bc}$ and $\Xi_{bb}$ according
to relevant publications. We name it as GENXICC and we test it by
comparing its numerical results with those in references. It is
written in a PYTHIA-compatible format and it can be easily
implemented into PYTHIA. GENXICC is also written in modularization
manner, with {\bf make}, a GNU C compiler, one may apply
the generator to various situations or experimental environments
very conveniently. \\


\end{abstract}

\maketitle

\noindent {\bf Program summary}\\

\noindent {\it Title of program}: GENXICC\\

\noindent {\it Version}: 1.0 (January, 2007)\\

\noindent {\it Program obtained from}:
http://www.itp.ac.cn/$\sim$zhangzx; CPC Program Library\\

\noindent {\it Computer}: Any LINUX based on PC with
FORTRAN 77 or FORTRAN90 and GNU C compiler being installed\\

\noindent {\it Operating systems}: LINUX\\

\noindent {\it Programming language used}: FORTRAN 77/90\\

\noindent {\it Memory required to execute with typical data}: About
2.0 MB.\\

\noindent {\it Distribution format}: tar gzip file\\

\noindent {\it Nature of physical problem}: Hadronic production of a
double-heavy baryons: $\Xi_{cc}$, $\Xi_{bc}$ and $\Xi_{bb}$.\\

\noindent {\it Method of solution}:  The production of the
double-heavy baryons is realized by producing a binding double-heavy
diquark either $(QQ')[^3S_1]_{\bar{3},6}$ ($Q,Q'=b,c$) or
$(QQ')[^1S_0]_{\bar{3},6}$, which is in color anti-triplet $\bar{3}$
or color sextuplet $6$ and in $S$-wave triplet or singlet
configuration respectively, and then by absorbing a proper light
quark non-perturbatively. For the production of the various
double-heavy baryons $\Xi_{cc}$, $\Xi_{bc}$\footnote{In fact, there
are two kinds of states for $\Xi_{bc}$, i.e., one is that the inside
$b$ and $c$ are symmetric in `flavor space' and the other is that
$b$ and $c$ are anti-symmetric in `flavor space' similar to the case
for the baryons $\Lambda$ and $\Sigma^0$. Let us call them as
$\Xi_{1bc}$ for symmetric one and $\Xi_{2bc}$ for antisymmetric one
when we need to distinguish them. Due to the electromagnetic
interaction between the quarks for instance, the two kinds of states
may have different masses (degeneracy broken).} and $\Xi_{bb}$, the
`gluon-gluon fusion' mechanism, being the most important, is written
precisely in the generator, but two additional mechanisms, i.e. the
`gluon-charm collision' and the `charm-charm collision' ones, only
for $\Xi_{cc}$ ($\Xi^+_{cc}$ or $\Xi^{++}_{cc}$) are written.
Furthermore, all the mechanisms are treated consistently within the
general-mass flavor-number (GM-VFN) scheme. Specially, to deal with
the amplitude and in order to save CPU time as much as possible, the
`improved helicity-approach' is applied for the most complicated
gluon-gluon fusion mechanism. The code with a proper option can
generate weighted and un-weighted events accordingly as user's wish.
Moreover, an interface to PYTHIA is provided to meet ones' needs to
generate the `complete events' of $\Xi_{QQ'}$, i.e. to do the
`showers' of the partons appearing in the initial and final states
of the subprocess, and the
hadronization for final obtained `showers' etc. \\

\noindent {\it Restrictions on the complexity of the problem}: In
GENXICC, the approach to the hadronic production in terms of a
`complete $\alpha_s^4$ calculation' via the production of a binding
diquark state either $(QQ)[^3S_1]_{\bar{3}}$ or $(QQ)[^1S_0]_6$
($Q=c, b$) for $\Xi_{cc}$ and $\Xi_{bb}$ production, and via that of
a binding diquark state of $(bc)[^3S_1]_{\bar{3}}$ or
$(bc)[^1S_0]_{\bar{3}}$ or $(bc)[^3S_1]_6$ or $(bc)[^1S_0]_6$ for
$\Xi_{bc}$ is available, but the contributions from the other higher
Fock states of the diquark states are not involved. Considering the
needs of comparisons and applications in most cases, three
mechanisms and their consistent summation for the hadronic
production of $\Xi_{cc}$ are available. But for most purposes and
applications to the baryons $\Xi_{bb}$ and $\Xi_{bc}$, which contain
$b$-quark(s) (much heavier than $c$-quark), only the `gluon-gluon
fusion' mechanism for the production is accurate enough, therefore,
here only the `gluon-gluon fusion' mechanism is available. Moreover,
since the polarization of the double-heavy baryons is also strongly
effected by hadronization of the double-heavy diquark produced via
the mechanisms considered here, so in the present generator only the
un-polarized production for the baryons are available. \\

\noindent{\it Typical running time}: It depends on which option one
chooses to match PYTHIA when generating the events and also on which
mechanism is chosen for generating the events. Typically, for the
most complicated case via gluon-gluon mechanism to generate the
mixed events via the intermediate diquark in $(cc)[^3S_1]_{\bar{3}}$
and $(cc)[^1S_0]_6$ states, then on a 1.8 GHz Intel P4-processor
PC-machine, if taking IDWTUP=1 for PYTHIA option (the meaning will
be explained later on), it takes about 20 hours to generate 1000
events, whereas, if IDWTUP=3 (the meaning will be explained later
on), it takes only about 40 minutes to generate $10^6$ events. \\


\section{introduction}

SELEX Collaboration has reported their observation of the doubly
charmed baryon $\Xi^+_{cc}$ \cite{exp,exp2} and their measured decay
width and production rate are much larger than those predicted
theoretically, e.g. Refs. \cite{baranov,kiselev1}. Nevertheless, it
started a new stage of the double-heavy baryon study. The
observations of the double-heavy baryons $\Xi_{QQ'}$,
$Q,Q'=b,c$\footnote{Throughout the paper, $\Xi_{QQ'}$ denotes the
baryons of $(QQ'q)$, $q=u,d$ a light quark, and the isospin-breaking
is ignored. For instance, $\Xi_{cc}$ denotes either $\Xi^+_{cc}$ or
$\Xi^{++}_{cc}$ etc.} should be done in all possible experimental
equipments, e.g. the detectors at the hadronic colliders (TEVATRON
and LHC) and fixed target ones as well etc. Especially, for the
lightest one $\Xi_{cc}$ so great discrepancy between the
experimental data and the theoretical predications should be
explained. Theoretically in order to explain the discrepacy, more
configurations of the double-heavy diquark inside the baryons and
more mechanisms, in addition to the gluon-gluon fusion mechanism,
have been taken into account\cite{cqww,cmqw}. Considering the usages
of experimental feasibility studies and the fact that the efficiency
to generate the double-heavy baryon $\Xi_{QQ'}$ with PYTHIA
\cite{pythia} directly is too low due to the reason as that in
generating $B_c$ meson, here in a PYTHIA-compatible format we write
an effective generator GENXICC for the hadronic production of
double-heavy baryons involving all of the formulae appearing in
Refs.\cite{cqww,cmqw} and extended to the production for the
double-heavy baryons $\Xi_{bc}$ and $\Xi_{bb}$ (heavier than
$\Xi_{cc}$). Especially, the double-heavy baryon $\Xi_{bc}$ has
quite large branching ratio for decaying into $J/\psi$ inclusively,
it may contribute serious background in observing $B_c$
experimentally. Since the format, GENXICC can be easily implemented
into PYTHIA to simulate the full events for various experimental
environments such as in the case of BCVEGPY for the $B_c$ meson
hadroproduction \cite{bcvegpy1,bcvegpy2,bcvegpy3}. Though the
double-heavy baryons $\Xi_{bc}$ and $\Xi_{bb}$ (heavier than
$\Xi_{cc}$) in the generator are involved, the cross-sections of
$\Xi_{bc}$ and $\Xi_{bb}$  are much smaller than that of $\Xi_{cc}$
and they are more difficult to be observed experimentally than
$\Xi_{cc}$. To meet the most needs in cases and to make the
generator GENXICC as compact as possible, here we only involve the
most important (great) gluon-gluon fusion mechanism for the
$\Xi_{bc}$ and $\Xi_{bb}$ production.

The formulation for the production in Refs.\cite{cqww,cmqw} is under
NRQCD framework\cite{nrqcd}. For instance, the $\Xi_{cc}$-baryon
production is realized via production of a binding diquark of heavy
flavors (a diquark core) either $(cc)[^3S_1]_{\bar{3}}$ (in
configuration of $S$-wave and in color anti-triplet $\bar{3}$) or
$(cc)[^1S_0]_6$ (in configuration of $S$-wave and in color sextuplet
$6$), and then the relevant double-heavy baryon is formed by means
of the diquark as a core to absorb a proper light quark from
`vacuum' non-perturbatively. Since there is no precisely theoretical
estimate on the relevant non-perturbative matrix element which
describes the fact of combining two heavy quarks into the diquark
and absorbing a proper quark from `vacuum' to form the baryon
non-perturbatively, thus as in
Refs.\cite{baranov,kiselev1,cqww,cmqw}, the matrix element here is
approximately taken as a `decay constant' of the heavy diquark core
and the probability of the diquark to form the baryon is assumed to
be one. Namely to study the hadronic production of $\Xi_{QQ'}$ is
equivalent to study the hadronic production of $(QQ')$-diquark. It
may be understood as that the fragmentation function $D(z)$ of a
heavy diquark into a baryon has a very sharp peak near $z=1$
\cite{kiselev1}, and the momentum of the final baryon may be
considered roughly equal to the momentum of the produced diquark.

In the generator, for $\Xi_{cc}$ production we include the
gluon-gluon fusion mechanism that via the subprocess
$g+g\to\Xi_{cc}+\bar{c}+\bar{c}$ and the extrinsic charm mechanisms
those via the subprocess $g+c\to\Xi_{cc}+\bar{c}$ (gluon-charm
collision) or the subprocess $c+c\to\Xi_{cc}+g$ (charm-charm
fusion), furthermore, two configurations of the binding diquark,
$(cc)[^3S_1]_{\bar{3}}$ and $(cc)[^1S_0]_{6}$, are considered as in
Refs.\cite{cqww,cmqw}. In the mentioned mechanisms, the heavy quark
components in the initial hadrons are those created from `gluon
splitting' perturbatively, so we call them as `extrinsic' ones for
convenience. Because it contains certain non-perturbative nature
\cite{cqww}, the extrinsic charm fusion via the subprocess $c+c\to
\Xi_{cc}$ is not included. There are also non-perturbative intrinsic
charm components in the initial hadrons \cite{brodsky}, which may
give sizable contributions when the C.M. energy of the collision is
not too great \cite{cmqw}. Since the contributions from the
`intrinsic charm' mechanism are model-dependent, i.e., they depend
on the non-perturbative intrinsic charm distribution function, so in
the generator we do not include the `intrinsic' mechanism.

The technology, which we will adopt here, for simulating the
production by taking the gluon-gluon fusion mechanism and the
extrinsic charm mechanism into account is that in Ref.\cite{cqww}.
Namely these mechanisms are consistently dealt with under the
general-mass flavor-number (GM-VFN) scheme \cite{gmvfn1,gmvfn2}.
Since here we will take the way to deal with the most complicated
mechanism of gluon-gluon fusion as that of the generator BCVEGPY,
which has been described quite well in
Refs.\cite{bcvegpy1,bcvegpy2,bcvegpy3}, so to shorten the paper, we
will not repeat it here (the interesting reader may consult
Ref.\cite{cqww}), but we will highlight only the differences at due
places.

Due to the fact that $b$-quark is much heavier than the $c$-quark,
to produce a $b\bar{b}$ pair is more difficult than to produce a
$c\bar{c}$ pair. Therefore, the production of $\Xi_{bb}$ and
$\Xi_{bc}$, in contrary to $\Xi_{cc}$, via extrinsic charm and/or
bottom mechanisms is much less important in comparison to that via
the gluon-gluon fusion mechanism. So considering the accuracy needed
in the most usages for simulating the double-heavy baryons
$\Xi_{bb}$ and $\Xi_{bc}$, we think that it is enough to consider
the gluon-gluon fusion mechanism but not the extrinsic charm and/or
bottom mechanisms, therefore, in the generator GENXICC, only the
gluon-gluon fusion mechanism is involved. The GENXICC package is
written in the format of PYTHIA\cite{pythia}, so the generator could
be easily implemented into PYTHIA. When the generator GENXICC is
implemented in PYTHIA, one can generate complete events of the
production conveniently as long as the created quark(s) and gluon in
the hard subprocess are connected to the relevant subprogram for
their fragmentation in PYTHIA. For user's convenience, we also set
up a switch in GENXICC to decide whether to use the VEGAS package
\cite{vegas} or not, that depends on whether user would like to
increase the Monte Carlo (MC) simulation efficiency for high
dimensional phase-space integration with VEGAS functions or not. If
using the VEGAS package, one may obtain the sampling importance
function that is necessary for the importance sampling MC method.

The paper is organized as follows. In Sec.II, we outline the
generator GENXICC, which includes an explanation of the generator
structure, the use of generator and some checks of the generator.
The final section is reserved for a summary.

\section{The generator: GENXICC}

Program GENXICC is a specific generator for hadronic production of a
baryon $\Xi_{cc}$, $\Xi_{bc}$ and $\Xi_{bb}$ in a Fortran package
form. Since the hadronic production of $\Xi_{bb}$ and $\Xi_{bc}$ are
similar to the case of $\Xi_{cc}$, in the paper we take the hadronic
production of $\Xi_{cc}$ as the main clue to explain the generator
GENXICC, and then share some lights on the production of $\Xi_{bc}$
and $\Xi_{bb}$ in suitable places.

\begin{figure}
\centering
\includegraphics[width=0.70\textwidth]{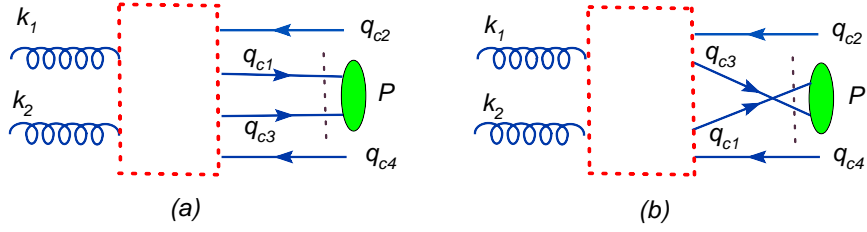}
\caption{The schematic Feynman diagrams for the hadroproduction of
$\Xi_{cc}$ from the gluon-gluon mechanism, where the dashed box
stands for the hard interaction kernel. $k_1$ and $k_2$ are two
momenta for the initial gluons, $q_{c2}$ and $q_{c4}$ are the
momenta for the two outgoing $\bar{c}$, $P$ is the momentum of
$\Xi_{cc}$. The $(cc)$-diquark pair is either in
$(cc)_{\bf\bar{3}}[^3S_1]$ or in $(cc)_{\bf 6}[^1S_0]$
respectively.} \label{feyn1}
\end{figure}

\begin{figure}
\centering \setlength{\unitlength}{1mm}
\begin{picture}(80,60)(30,30)
\put(-37,-110) {\includegraphics{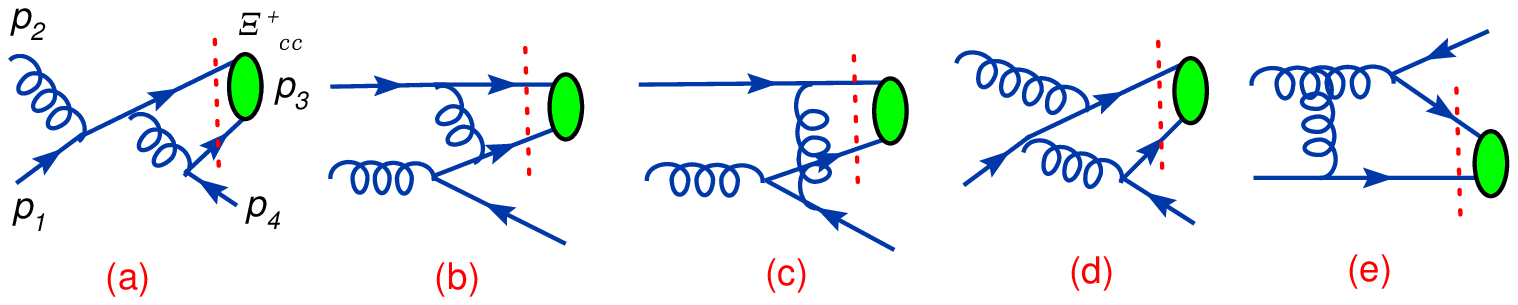}}
\put(-37,-140) {\includegraphics{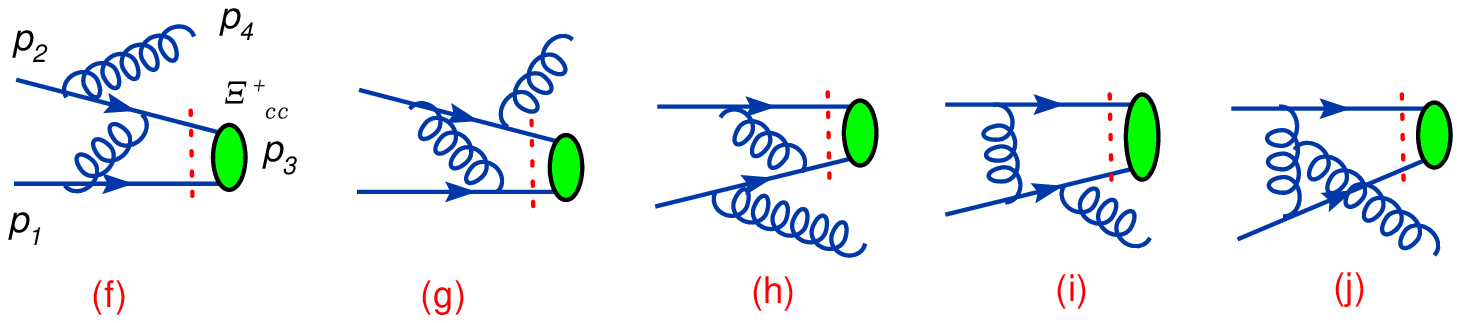}}
\end{picture}
\caption{Typical Feynman diagrams for the sub-processes induced by
`extrinsic' charm. The upper five of them are those for
$g(p_{1})+c(p_{2})\to \Xi_{cc}(p_{3})+\bar{c}(p_{4})$ and the lower
five are for $c(p_{1})+c(p_{2})\to \Xi_{cc}(p_{3})+g(p_{4})$
respectively. The $(cc)$-diquark pair is either in
$(cc)_{\bf\bar{3}}[^3S_1]$ or in $(cc)_{\bf 6}[^1S_0]$
respectively.} \label{feyn2}
\end{figure}

To produce $\Xi_{cc}$ by hadronic collisions, the comparatively
important mechanisms have: the gluon-gluon fusion one via the
subprocess $g+g\to\Xi_{cc}+\bar{c}+\bar{c}$ and the extrinsic charm
one via the subprocess of the gluon-charm collision
$g+c\to\Xi_{cc}+\bar{c}$ or the charm-charm collision with a hard
gluon emission $c+c\to\Xi_{cc}+g$. The schematic Feynman diagrams
for the gluon-gluon fusion mechanism are shown in Fig.(\ref{feyn1}).
Fig.(\ref{feyn1}) shows the two ways to form the $(cc)$-diquark pair
from the outgoing valence quarks, and each way simply contains 36
Feynman diagrams that are similar to the case of hadronic production
of $B_c$ (all the diagrams can be found in Ref.\cite{bcvegpy1}, but
one need to change all the $b$ quark line there to the $c$ quark
line). For each of the sub-processes $g+c\to \Xi_{cc}+\bar{c}$ and
$c+\bar{c}\to \Xi_{cc}+g$ of the extrinsic charm mechanism, there
are ten Feynman diagrams, and the typical Feynman diagrams of them
are shown in Fig.(\ref{feyn2}). Moreover, for every mechanism the
intermediate diquark may be in the configurations either
$(cc)[^3S_1]_{\bf\bar{3}}$ or $(cc)[^1S_0]_{\bf 6}$. In the
generator, the program is written as that the events are generated
by specific mechanism and configuration of the double-heavy diquark
separately, whereas if it is needed to simulate the true situation,
the events generated by these distinguishable mechanisms and diquark
configurations can be 'mixed' by taking proper options provided by
PYTHIA. More explanation about generating mixing events with PYTHIA
can be found in PYTHIA mannual \cite{pythia}.

Since GENXICC is written in the format as that of PYTHIA (including
common block variables), so all of the mentioned mechanisms and
double-heavy diquark configurations can be easily implemented into
PYTHIA as an external process according to user's specific wish.
Thus all the functions in PYTHIA can be applied to GENXICC
conveniently. We should note here that one has to set the PYTHIA
parameter ${\rm mstp(61)} \equiv 0$ when an extrinsic massive charm
mechanism is considered, that means the showers for the initial
state particles especially for that of the massive $c$-quark are
switched off. It is because that the present version of PYTHIA can
deal with the initial state showers only for massless
particles\footnote{In PYTHIA, one way to solve the problem is
introduced by transforming the initial massive particle to an
equivalent massless one \cite{pythia}, but it fails for the present
case.}.

GENXICC may be more conveniently used to simulate the events
according to various experimental situations with a GNU C compiler
under the LINUX environment at a PC than anyone under the WINDOWS
operating system. Thus we suggest that one had better to work with
GENXICC under a LINUX environment with FORTRAN 90 (or FORTRAN 77)
and GNU C compiler being installed. The command {\bf make} in GNU is
used to arrange the whole execution of the generator. When running
the GNU command {\bf make}, firstly all the C/C++ codes in the
configuration file {\bf makefile}, which is in the main directory,
are compiled sequentially, and all of the FORTRAN source files of
the program are followed to be also compiled, then all of the
compiled files are linked according to these C/C++ codes
automatically, finally an executable file with the default name {\bf
run} will be built. By running the built executable file {\bf run},
one can generate the double-heavy baryon $\Xi_{cc}$ ($\Xi_{bc},
\Xi_{bb}$) events accordingly. Under this way, the time for
compiling the Fortran source files is saved a lot. This point will
be explained more in the following subsection. We note here that all
the sources files in our package are written in FORTRAN only i.e.
its C/C++ version is not available so far, while as described above,
alternatively one may use the GNU C compiler {\bf make} to run the
generator with FORTRAN source files to meet all kinds of needs
instead. Furthermore, a simple script, named as {\bf do}, which does
all the necessary jobs for generating the events, e.g. calling the
command {\bf make} to generate the executable file {\bf run} and
running the builded executable file {\bf run} and etc. is put in the
main directory of the program.

We also note that the polarization of the production for the
double-heavy baryons has not been programmed. In fact, to estimate
the polarization of the produced double-heavy baryons is much more
complicated problem than that of unpolarized one. At this moment, in
the program we, also as the authors in literature, treat the
production in such a way just according to NRQCD factorization: if
taking $\Xi_{cc}$ as an example, firstly a proper double-heavy
diquark $(cc)$ in $^1S_0$ or $^3S_1$ configuration is produced, that
can be computed by pQCD, then the diquark to form the baryons in
terms of non-perturbative QCD, that is not calculable and is
dictated by the so-called `matrix element'. Furthermore as that in
literature to deduce the input parameters as possible, the
polarization of the baryon is summed up in the program for the
production via the diquark and spin-symmetry for the matrix element
is assumed, that the $^3S_1$ diquark configuration to produce a
double-heavy baryon in spin-half or spin three halves by capture a
light quark from vacuum is `averaged' respect to spin of the binding
diquark. Therefore, there is a weight factor, $\frac{2}{3}$, for
producing the double-heavy $\Xi^*_{cc}$ in spin three halves and a
weight factor, $\frac{1}{3}$, for producing the double-heavy
$\Xi_{cc}$ in spin a half with the assumption. Since the excited
state $\Xi^*_{cc}$ (spin in three halves) will decay into the ground
state (spin in one half) with an almost 100\% possibility via strong
and/or electromagnetic interaction, so considering `comparatively
poor' abilities of experiments in the foreseen future and for
simplifying the problem, we consider the production of $\Xi^*_{cc}$
and $\Xi_{cc}$ together, so that without distinguishing the two
states the weight factors of them should be added (to become $1$).
In this sense later on, if there is no special emphasis, the
production of $\Xi_{cc}$ means the production of $\Xi_{cc}$ and
$\Xi^*_{cc}$ as well always so the matrix elements $h^{cc}_3$ and
$h^{cc}_1$ mean the hadronization of the diquarks to the baryons
$\Xi_{cc}$ and $\Xi^*_{cc}$ as well. For the production of
$\Xi_{bb}$ and $\Xi_{bc}$ here the treatments are the same. Even so,
for the production concerned, we still need to know the values of
the matrix elements $h^{cc}_3, h^{cc}_1, h^{bb}_3, h^{bb}_1,
h^{bc}_3, h'^{bc}_3, h^{bc}_1$ and $h'^{bc}_1$ due to the various
intermediate double-heavy diquarks. If someone is interested in
polarization of the $^3S_1$ configuration specially, only for the
gluon-gluon fusion mechanism he can realize his interest by making
proper changes in GENXICC, i.e., to make suitable changes in the
subroutine {\bf bundhelicity} (in the file s$\_$bound.f of the
subdirectory {\it ggsub}) accordingly, he can generate the diquark
with a definite polarization as he likes. More explicitly, for
GENXICC in default status, summation over the three components of
the $^3S_1$ diquark's polarization is programmed in the subroutine
{\bf bundhelicity}, therefore, the one can achieve the polarized
result by simply keeping the terms corresponding to the component of
the diquark's polarization of his interest, and commenting out the
remaining terms related to his not interest components of the
diquark's polarization in {\bf bundhelicity}.

\subsection{Structure of the generator}

\begin{figure}
\centering
\includegraphics[width=0.70\textwidth]{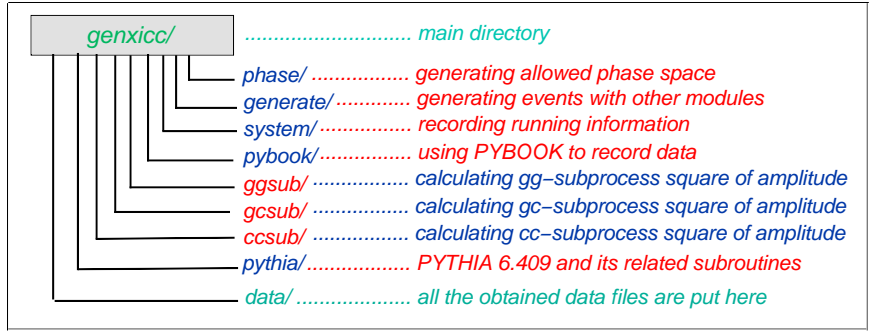}
\caption{The schematic structure of GENXICC.} \label{struct}
\end{figure}

The schematic structure of the program is shown in FIG.\ref{struct}.
It shows that there are totally nine subdirectories in the main
directory {\it genxicc}. The subdirectory {\it data} is used to
store all the obtained data files, while the other eight
subdirectories contain necessary files to complete specific tasks
accordingly for generating events. Namely the generator GENXICC has
been divided into eight modules with less cross communication among
the modules, and each module is applied to complete assigned
task(s). In the main directory {\it genxicc}, there are three
Fortran source files: parameter.F, run.F and xicc.F. The file xicc.F
is the main program of the generator, which arranges the running of
all the Fortran codes in the main directory and in subdirectories in
proper order. The file run.F is to define a few of global variables
(all of which are named as the variables in capital, e.g.
ENERGYOFLHC (the collision energy for LHC) and NUMOFEVENTS (the
number of events to be generated) etc. The file parameter.F is to
set the initial values for all the other parameters, such as the
mass of $c$-quark etc.. The eight modules of GENXICC include:

\begin{itemize}

\item The module {\bf generate}: it is the key module of the program
and it contains the files for generating $\Xi_{cc}$, $\Xi_{bc}$ and
$\Xi_{bb}$ events respectively with the help of the modules {\bf
phase}, {\bf ggsub}, {\bf gcsub}, {\bf ccsub} and {\bf pythia}. The
functions of the module {\bf generate} are to set the initialize
condition for event simulation; to establish the connection with
PYTHIA \cite{pythia} and a connection to a specific version of the
parton distribution functions (PDFs) that have not been included in
PYTHIA yet but included in the package for convenience (i.e. the
three types of PDFs: CTEQ6HQ \cite{6hqcteq}, GRV98L \cite{98lgrv}
and MRST2001L \cite{2001lmrst}); to calculate the integrand (the
squared amplitude with necessary Jacobian for the phase space
integration) with the help of either the module {\bf ggsub} for
gluon-gluon fusion mechanism or the module {\bf gcsub} for gluon
extrinsic-charm mechanism or the module {\bf ccsub} for gluon
extrinsic-bottom mechanism; to do the phase space integration with
the help of the module {\bf phase}. The module {\bf generate}
contains five Fortran source files: evntinit.F, outerpdf.F,
totfun.F, sepfun.F and mixfun.F. Particularly, for simulating
$\Xi_{QQ'}$ production with a given mechanism, sepfun.F is to
calculate the integrand (the squared absolute amplitude with Jacobi
determinant of the phase integration) for a certain $(QQ')$-diquark
production channel as specified by the parameter {\bf ixiccstate}
(the meaning of which will be explained in the next subsection);
mixfun.F is to calculate the integrand for the summed results, e.g.
to calculate the summed results of the $(QQ')$-diquark in possible
spin-color configurations for a certain mechanism as specified by
the parameter {\bf imixtype} (the meaning of which will be explained
in the next subsection).

\item The module {\bf phase}: it contains the files for generating
the allowed phase-space point and for generating an importance
sampling function with the help of the VEGAS program \cite{vegas}.
There are four source files: phase$\_$gen.F, phase$\_$point.F,
charmsub.F and vegas.F. The file phase$\_$point.F is to generate the
phase-space points allowed kinematically with the help of the file
phase$\_$gen.F; the file phase$\_$gen.F contains the RAMBO
subroutine \cite{rambos}, which can be used to generate the allowed
phase space and to generate the momenta of the particles in final
state; the file charmsub.F is only used in the case of $\Xi_{cc}$
production for calculating the subtraction terms of the charm PDF,
that is necessary to avoid the `double counting' when summing up the
results of the two mechanisms: the gluon-gluon fusion and the
gluon-charm collision \cite{cqww}; the file vegas.F contains the
VEGAS program, which is used to generate the importance sampling
function for MC simulation.

\item The module {\bf ggsub}: it contains the files for
calculating the squared amplitudes of $\Xi_{cc}$, $\Xi_{bb}$ and
$\Xi_{bc}$ production through the gluon-gluon fusion subprocess. As
for $\Xi_{cc}$ or $\Xi_{bb}$ simulation, when computing the squared
amplitudes, the intermediate diquark is in either one of the two
spin-color configurations $(QQ)[^3S_1]_{\bar{3}}$ and
$(QQ)[^1S_0]_6\, (Q=b,c)$ respectively. While for $\Xi_{bc}$, the
intermediate diquark can be in either one of the four spin-color
configurations $(bc)[^{3}S_{1}]_{\bar{3}}$, $(bc)[^{3}S_{1}]_{6}$,
$(bc)[^{1}S_{0}]_{\bar{3}}$ and $(bc)[^{1}S_{0}]_{6}$. There are
five Fortran source files: s$\_$bound.F, s$\_$common.F,
s$\_$foursets.F, s$\_$free.F and s$\_$samp.F. The files are taken
from BCVEGPY \cite{bcvegpy3}, but necessary changes are made so as
to make them suitable for the present purpose (i.e., to compute the
$\Xi_{cc}$, $\Xi_{bc}$ and $\Xi_{bb}$ production). Here due to the
replacement of an antiquark by a quark, the changes are included in
color factors, the connection between the free quark part amplitudes
and the bound state part amplitudes and etc.. More differences
between the production of $\Xi_{cc}$ ($\Xi_{bc}$ and $\Xi_{bb}$) and
the production of $B_c$ through the gluon-gluon fusion subprocess
can be found in Ref.\cite{cqww}.

\item The module {\bf gcsub}: it contains files for
calculating the amplitudes squared for $\Xi_{cc}$ production through
the gluon-charm collision (the gluon extrinsic charm mechanism).
When computing the amplitudes squared, and the intermediate
$(cc)$-diquark configuration either $(cc)[^{3}S_{1}]_{\bf \bar{3}}$
or $(cc)[^{1}S_{0}]_{\bf 6}$ is considered respectively. There are
two source files: gcamp1.F and gcamp2.F. The file gcamp1.F is to
calculate the amplitude squared for the configuration
$(cc)[^{3}S_{1}]_{\bf \bar{3}}$ and the file gcamp2.F is to
calculate the amplitude squared for the configuration
$(cc)[^{1}S_{0}]_{\bf 6}$.

\item The module {\bf ccsub}: it contains the files for
calculating the amplitudes squared for producing $\Xi_{cc}$ through
the charm-charm collision (via the subprocess $c+c\to\Xi_{cc}+g$).
When computing the amplitudes squared, the intermediate
$(cc)$-diquark configuration either $(cc)[^{3}S_{1}]_{\bf \bar{3}}$
or $(cc)[^{1}S_{0}]_{\bf 6}$ is considered respectively. There are
two source files: ccamp1.F and ccamp2.F. The file ccamp1.F is to
calculate the squared amplitude for the configuration
$(cc)[^{3}S_{1}]_{\bf \bar{3}}$ and the file ccamp2.F is to
calculate the amplitude squared for the configuration
$(cc)[^{1}S_{0}]_{\bf 6}$.

\item The module {\bf pythia}: it contains the files for PYTHIA running.
There are four source files: py6409.F (a nickname for PYTHIA 6.409),
upinit.F, upevnt.F and colorflow.F. Here upinit.F and upevnt.F are
two files to initialize {\bf an external process defined by user}
for PYTHIA. While colorflow.F sets the color flow information for
the concerned processes. For convenience, the manual
(pythia$\_$manual.tar.gz) and some notes (pythia6409.update) for the
current PYTHIA version 6.409 are included.

\item The module {\bf pybook}: it contains the files for
initializing the subroutine PYBOOK of PYTHIA to record the events.
The subroutine PYBOOK can be easily adapted to achieve the
differential distributions of a certain particle in final state.
Note here that one can record all the necessary information of the
events at one run by properly using of PYBOOK. Some typical ways to
record the events with PYBOOK have been programmed in the generator
GENXICC. Furthermore, the user may conveniently switch off this
module in xicc.F in main directory so as to use his/her own way to
record the data. There are five source files: pybookinit.F,
uphistrange.F, uppydump.F, uppyfact.F and uppyfill.F. Here, the four
files: pybookinit.F, uppydump.F, uppyfact.F and uppyfill.F are to
call the PYTHIA subroutines PYBOOK, PYDUMP, PYFACT, PYFILL
respectively; uphistrange.F is to set the ranges for the histograms.

\item The module {\bf system}: it contains the
files to open or to close the relevant record files and to print out
certain running messages at the intermediate steps according to
one's wish. They tell the user at which step the program is running.
There are seven source files: upopenfile.F, uplogo.F, vegaslogo.F,
updatafile.F, upclosegradefile.F, uperror.F and upclosepyfile.F.
Here upopenfile.F and upclosegradefile.F are used to open or to
close the files for recording the sampling importance function;
updatafile.F and upclosepyfile.F are used to open or to close the
data files used for PYBOOK; uperror.F is to print some possible
warning messages during program running; uplogo.F is to print the
logo of the generator to the screen or to the data file with suffix
`.cs'; vegaslogo.F is to print the running information during VEGAS
running to the screen or to the data file with suffix `.cs'.

\end{itemize}

All the files for recording the running information are put in the
subdirectory {\bf data}. To distinguish them, all the grade files
(sampling importance function) for recording the obtained data are
ended with the suffix `.grid', all the intermediate files, which
record the used parameter values and the VEGAS running information,
are ended with the suffix `.cs' and all the files which record the
differential distributions, e.g. the transverse momentum and
rapidity of a specific particle in final state, such as $\Xi_{cc}$,
are ended with the suffix `.dat'.

Each module is equipped with its own {\bf makefile} that will be
used to make a library of the same name, e.g. the {\bf makefile} in
the subdirectory {\bf generate} will be used by the GNU command {\bf
make} to generate a library named as {\it generate.a}, which is
located in the main directory. These sub-makefiles are orchestrated
by a master {\bf makefile} in the main directory. Under this way,
the time for compiling Fortran source files can be saved a lot,
because once a source file is compiled, one does not need to
re-compile it again, unless some changes are made. Libraries
required for the main program are listed in the LIBS variable of the
master {\bf makefile} and built automatically by invoking the
sub-makefiles:
\begin{displaymath}
{\rm LIBS} = generate.a\;\; phase.a\;\; system.a\;\; pybook.a\;\;
pythia.a\;\; ggsub.a\;\; gcsub.a\;\; ccsub.a
\end{displaymath}

Under the way based on {\bf makefile}, with the command {\bf make}
an executable file is built whose default name is {\bf run}, then
the program acquires good modularity and re-usability. Thus, the
user can easily reform the generator to suit his/her experimental
environment. A simple script, which is named as {\bf do} and does
all the necessary jobs for generating events, is put in the main
directory. For convenience, another script, taken directly from the
FormCalc package \cite{formcalc}, is also supplied, which is named
as {\bf pnuglot} and may produce a high-quality Encapsulated
PostScript (EPS) file to plot the generated data in relevant file(s)
that is(are) in the subdirectory {\bf data}.

\subsection{Flow charts of generator}

\begin{figure}
\centering
\includegraphics[width=0.40\textwidth]{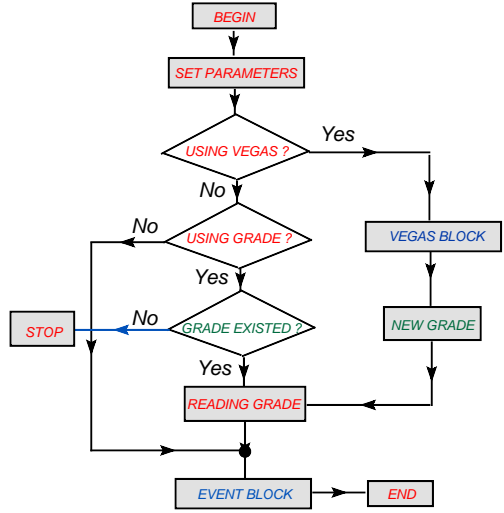}
\caption{The overall schematic flow chart of the generator GENXICC.}
\label{chart1}
\end{figure}

\begin{figure}
\centering
\includegraphics[width=0.40\textwidth]{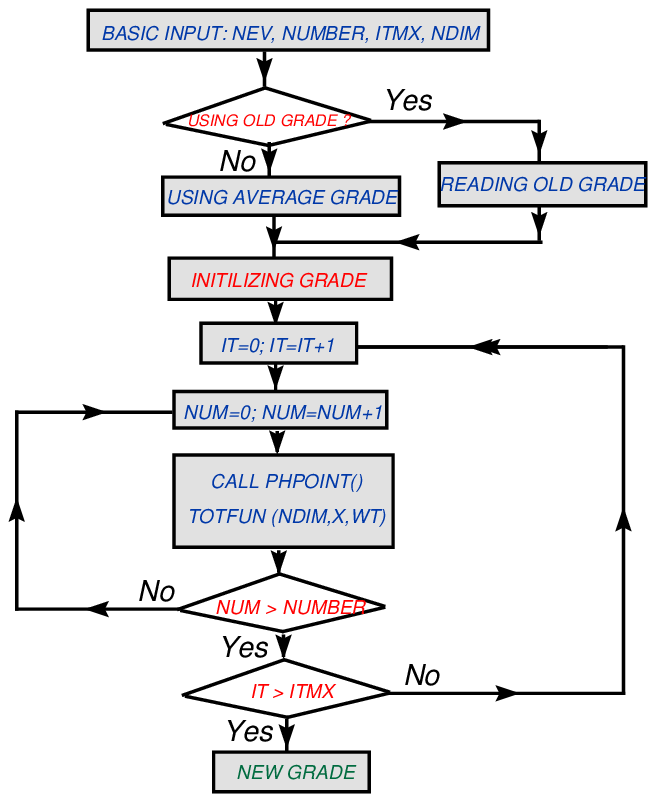}\hspace{1.5cm}
\includegraphics[width=0.34\textwidth]{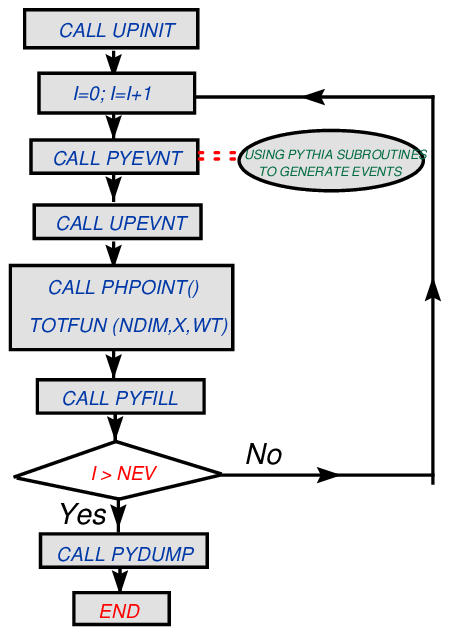}
\caption{The schematic flow charts for the {\it vegas} block (on the
left) and the {\it event} block (on the right) of the generator
GENXICC.} \label{chart2}
\end{figure}

The overall schematic flow chart of the generator GENXICC is shown
in Fig.(\ref{chart1}). It shows that GENXICC can be schematically
divided into two blocks, i.e. the {\it vegas} block (mainly with
{\bf phase}) and the {\it event} block (mainly with {\bf generate}).
The function of the {\it vegas} block is to generate the sampling
importance function that is necessary for importance sampling, and
the function of the {\it event} block is to generate events with
PYTHIA, e.g. all the mentioned production processes for the hadornic
production of $\Xi_{QQ'}$ can be implemented straightforward in
PYTHIA as external process by properly programming the two PYTHIA
subroutines UPINIT and UPEVNT \footnote{The UPINIT and UPEVNT are
two necessary subroutines to define external process for PYTHIA:
UPINIT is called by PYTHIA subroutine PYINIT and is used to set the
initial conditions for the user-defined process, and UPEVNT is
called by PYTHIA subroutine PYEVNT and is used to set the event
conditions of the user-defined process.}. The schematic flow charts
for the {\it vegas} block and the {\it event} block of the generator
GENXICC are shown in Fig.(\ref{chart2}).

When simulating the production, one may chose whether to take
importance sampling method with VEGAS or not. If the importance
sampling method is used the simulation efficiency may increase a
lot. The VEGAS package is programmed to achieve a sampling
importance function. One can learn from Fig.(\ref{chart1}) that
there are three ways to do the event simulation in the program: one
is the trivial MC and the left two ways are to use the importance
sampling method, i.e. one is to use the existent grades (importance
sampling function) that have been generated in previous run(s) and
have been recorded in .grid file(s) in the subdirectory {\bf data}
and the other one is to use a new grade generated by the current
VEGAS running. When setting IVEGASOPEN=0 and IGRADE=1 for instance
when the function has been recorded in .grid file(s) by previous
run(s) already, i.e. adopting the first importance sampling method,
one may generate $\Xi_{cc}$ events, just by reading the existent
importance sampling function. Thus, when using the package under
these proper options, one just needs to run VEGAS once enough.

The precision of the generated importance sampling function can be
improved by properly adjusting the maximum iteration number, the
number of calls to the integrand in each time of the iteration, and
the number of bins (the $[0,1]$ region is divided into how many
sub-regions) as well. For convenience, we define three parameters in
the head file invegas.h: NVEGITMX (maximum number of allowed
iterations), NVEGCALL (maximum total number of the times to call the
integrand in each iteration) and NVEGBIN (the number of bins). A
subtle point is that, in VEGAS, the default number of NVEGBIN is 50
(i.e. the region of (0,1) is divided into 50 pieces); by taking a
bigger proper value for NVEGBIN, one can improve the production
efficiency but need more iterations to obtain a stable result. In
practice, the values for NVEGITMX, NVEGCALL and NVEGBIN should be
carefully chosen to obtain the best important sampling function
within the least of cpu time. In our program, we take the default
value of the column bins to be 100 (according to our experience, it
is enough); and the user who wants to do some very precise studies,
a proper variation of NVEGBIN is needed. And in practice, we have
found that for the most complex gluon-gluon fusion mechanism for the
hadronic production of $\Xi_{cc}$, $\Xi_{bb}$ and $\Xi_{bc}$, if
choosing NVEGITMX=20 and NVEGCALL=300000, then we can achieve the
precision goal of $0.1\%$ for the total cross-section.

The flow chart for the {\it event} block is shown in the right
diagram of Fig.(\ref{chart2}). It is to generate events with the
help of PYTHIA package. At the present, we adopt the PYBOOK
subroutines that are provided by PYTHIA, such as PYFILL, PYDUMP,
PYFACT and etc., to record all the obtained information of the
events. However, users may comment out all the calls for PYBOOK
subroutines in the source file xicc.F and then add his/her own ways
to record the events conveniently.

\subsection{Use of the generator}

GENXICC can generate a huge event sample of producing $\Xi_{cc}$ or
$\Xi_{bc}$ or $\Xi_{bb}$ efficiently in the case of hadronic
collisions.  As for the hadronic production of $\Xi_{QQ}$ ($Q=c,
b$), in framework of NRQCD \cite{nrqcd}, the partonic cross-section
of the concerned mechanisms can be expressed as \cite{majp,cmqw}:
\begin{equation}
\hat{\sigma}_{ab\rightarrow \Xi_{QQ}} = H (ab\to
(QQ)[^3S_1]_{\bf\bar 3} ) \cdot h^{QQ}_3 +H (ab\to (QQ)[^1S_0]_{\bf
6} ) \cdot h^{QQ}_1 +\cdots, \label{nrqcd10}
\end{equation}
where the ellipsis stands for the terms in higher orders of $v$. $H
(ab\to (QQ)[^3S_1]_{\bf\bar 3} )$ or $ H (ab\to (QQ)[^1S_0]_{\bf 6}
)$ is the perturbative coefficient for producing a $QQ$ pair in
configuration of $^3 S_1$ and color ${\bf\bar 3}$, or in
configuration of $^1 S_0$ and color ${\bf 6}$ respectively. The
parameters $h^{QQ}_3$ and $h^{QQ}_1$ are the values of the
nonperturbative `matrix elements' in NRQCD to characterize the
transitions of the charm pair of quarks in $[^3S_1]_{\bf\bar 3}$ and
in $[^1S_0]_{\bf 6}$ into the baryon $\Xi_{QQ}$ respectively.
According to the argument presented in the INTRODUCTION, the
hadronic production of $\Xi_{QQ}$ is `equivalent' to the hadronic
production of $(QQ)$-diquark. Under such condition, the value of
NRQCD matrix element $h^{QQ}_3$ can be related to the wavefunction
for the color anti-triplet $[^3S_1]$ $(QQ)$-diquark state, i.e.
$h^{QQ}_3=\delta\cdot|R_{QQ}(0)|^2/4\pi$, where $R_{QQ}(0)$ is the
wave function at origin for the diquark binding system $(QQ)$ and
$\delta$ is the weight factor from `polarization average assumption'
in producing the baryon that is mentioned above). While as for
$\Xi_{bc}$, the factorization formula is different from that for
$\Xi_{QQ}$ due to different flavors being involved and, in fact,
there are two physical states $\Xi_{1bc}$ and $\Xi_{2bc}$, thus it
reads:
\begin{eqnarray}
&\hat{\sigma}_{ab\rightarrow \Xi_{1bc}} = H (ab\to
(bc)[^3S_1]_{\bf\bar 3} ) \cdot h^{bc}_3+H (ab\to (bc)[^1S_0]_{\bf
6} ) \cdot h^{bc}_1 +\cdots \nonumber\\
&\hat{\sigma}_{ab\rightarrow \Xi_{2bc}} = H (ab\to
(bc)[^1S_0]_{\bf\bar 3} ) \cdot {h'}^{bc}_3 +H (ab\to
(bc)[^3S_1]_{\bf 6} ) \cdot {h'}^{bc}_1 +\cdots . \label{nrqcd101}
\end{eqnarray}
More explicitly, four matrix elements: $h^{bc}_1$, ${h'}^{bc}_1$,
$h^{bc}_3$ and ${h'}^{bc}_3$, which characterize the transitions of
the bottom-charm diquarks in $[^1S_0]_{\bf 6}$, $[^3S_1]_{\bf 6}$,
$[^3S_1]_{\bf\bar 3}$ and $[^1S_0]_{\bf\bar 3}$ into $\Xi_{bc}$
respectively are needed. Similarly $h^{bc}_3$ can be related to the
wavefunction of the color anti-triplet diquark $(bc)[^3S_1]$ as
$h^{bc}_3=\delta\cdot |R_{bc}(0)|^2/4\pi$.

Users may communicate with or give instructions to the program
through the source files run.F and parameter.F. These two input
files allow users to setup the generation parameters and requests.
All the necessary parameters are:

\begin{itemize}

\item  pmc=: mass of $c$ quark (in units GeV). It should be taken as
an effective one that can be derived from the value of the pole mass
and the $\overline{MS}$ running mass \cite{ccfl}. Its default value
is taken to be $1.75$~GeV.

\item  pmb=: mass of $b$ quark (in units GeV). Similarly, it should be taken as
an effective one and its default value is taken to be $5.1$~GeV.

\begin{table}
\begin{center}
\caption{All the considered channels for the hadronic production of
the double-heavy baryon, which are defined by the two parameters
{\bf mgenxi} and {\bf ixiccstate}. Here the symbol gg-channgel
stands for the gluon-gluon fusion channel and etc..} \vskip 0.6cm
\begin{tabular}{|c||c|c|c|}
\hline ~~---~~ & ~~mgenxi=1 (for $\Xi_{cc}$)~~
& ~~mgenxi=2 (for $\Xi_{bc}$)~~ & ~~mgenxi=3 (for $\Xi_{bb}$)~~ \\
\hline\hline ~~ixiccstate=1~~ & gg-channel, $(cc)_{\bf
\bar{3}}(^3S_1)$ & gg-channel, $(bc)_{\bf \bar{3}}(^3S_1)$
& gg-channel, $(bb)_{\bf \bar{3}}(^3S_1)$ \\
\hline ~~ixiccstate=2~~ & gg-channel, $(cc)_{\bf 6}(^1S_0)$
& gg-channel, $(bc)_{\bf 6}(^1S_0)$ & gg-channel, $(bb)_{\bf 6}(^1S_0)$ \\
\hline ~~ixiccstate=3~~ & gc-channel, $(cc)_{\bf \bar{3}}(^3S_1)$
& gg-channel, $(bc)_{\bf 6}(^3S_1)$ & ~~---~~ \\
\hline ~~ixiccstate=4~~ & gc-channel, $(cc)_{\bf 6}(^1S_0)$
& gg-channel, $(bc)_{\bf \bar{3}}(^1S_0)$ & ~~---~~ \\
\hline ~~ixiccstate=5~~ & cc-channel, $(cc)_{\bf \bar{3}}(^3S_1)$
& ~~---~~ & ~~---~~ \\
\hline ~~ixiccstate=6~~ & cc-channel, $(cc)_{\bf 6}(^1S_0)$
& ~~---~~ & ~~---~~ \\
\hline
\end{tabular} \label{xi}
\end{center}
\end{table}

\item  mgenxi=: symbol to choose which double-heavy baron,
i.e. $\Xi_{cc}$, $\Xi_{bb}$ and $\Xi_{bc}$, to be generated.
mgenxi=1 is to generate $\Xi_{cc}$ events, mgenxi=2 is to generate
$\Xi_{bc}$ events and mgenxi=3 is to generate $\Xi_{bb}$ events. As
for the production of $\Xi_{bb}$ and $\Xi_{bc}$, only their dominant
gluon-gluon fusion mechanism is programmed. For clarity, we show all
the considered channels in TAB.\ref{xi}. In default, we set
mgenxi=1.

\item  pmxicc=: the mass of $\Xi_{cc}$, $\Xi_{bc}$ and $\Xi_{bb}$ (in units GeV)
respectively according to the value of mgenxi: pmxicc=$M_{\Xi_{cc}}$
for mgenxi=1; pmxicc=$M_{\Xi_{bc}}$ for mgenxi=2;
pmxicc=$M_{\Xi_{bb}}$ for mgenxi=3. To ensure the gauge invariance
of the hard scattering amplitude, we must set pmxicc=pmc+pmc for
$\Xi_{cc}$, pmxicc=pmb+pmc for $\Xi_{bc}$ and pmxicc=pmb+pmb for
$\Xi_{bb}$.

\item  fxicc=: the value of $R_{cc}(0)$ or
$R_{bc}(0)$ or $R_{bb}(0)$ (in units ${\rm GeV}^{3/2}$)
corresponding to the value of mgenxi: fxicc=$R_{cc}(0)$ for
mgenxi=1; fxicc=$R_{bc}(0)$ for mgenxi=2; fxicc=$R_{bb}(0)$ for
mgenxi=3. The default values for mgenxi=1,2,3 are taken as
$0.70$GeV$^{3/2}$, $0.90$GeV$^{3/2}$ and $1.38$GeV$^{3/2}$
\cite{baranov} respectively.

\item cmfactor: the relation among the non-perturbative matrix elements
as defined above. It has been pointed out in \cite{majp} that the
relevant matrix elements have the values at the same order of $v$.
So for convenience, we introduce an addition parameter ${\rm
cmfactor}$ which may vary in a possible region and further assume
that $h^{cc}_1\simeq {\rm cmfactor}\cdot h^{cc}_3$, $h^{bb}_1\simeq
{\rm cmfactor}\cdot h^{bb}_3$ and $h^{bc}_1\simeq {h'}^{bc}_1\simeq
{h'}^{bc}_3\simeq {\rm cmfactor}\cdot h^{bc}_3$. One may study
certain dependence of the production on the matrix elements by
setting the value of the parameter. Note that here we take ${\rm
cmfactor}=1.0$ as its default value.

\item  ptcut=:  $p_\mathrm{T}$ (in units GeV) cut for $\Xi_{cc}$,
$\Xi_{bc}$ and $\Xi_{bb}$ respectively. In default, we set
ptcut=0.2GeV.

\item  etacut=:  rapidity cut for $\Xi_{cc}$, $\Xi_{bc}$
and $\Xi_{bb}$ respectively. In default, there is no rapidity cut
for the production.

\item  psetacut=:  pseudo-rapidity cut for $\Xi_{cc}$, $\Xi_{bc}$
and $\Xi_{bb}$ respectively. In default, there is no pseudo-rapidity
cut for the production.

\item  npdfu=: choice of the collision type of
hadrons. npdfu=$1$ for $p-\bar{p}$ collision with ${\rm
ecm}=1.96$~TeV (TEVATRON), npdfu=$2$ for $p-p$ collision with ${\rm
ecm}=14.0$~TeV (LHC) and npdfu=$3$ for a fixed target experiment,
but in the program we take it such as $p-\bar{p}$ collision with
${\rm ecm}=33.56$~GeV (SELEX). In default, we set npdfu=1.

\item  ecm=: total energy for the hadron collision (in
units GeV). It is set to be ENERGYOFTEVA, ENERGYOFLHC or
ENERGYOFSELEX in accordance to the value of the parameter npdfu.

\item  imix=: whether to generate the mixed events.
imix=0 for generating separate events (e.g. events from a particular
mechanism and a fixed configuration of the intermediate diquark
state) according to the value of ixiccstate; imix=1 for generating
mixed events according to the value of imixtype. As for mgenxi=1, it
is to generate mixed events of $(cc)[^3S_1]_{\bf\bar 3}$ and
$(cc)[^1S_0]_{\bf 6}$ of the same mechanism. As for mgenxi=2, it is
to generate mixed events of $(bc)[^3S_1]_{\bf\bar 3}$,
$(bc)[^3S_1]_{\bf 6}$, $(bc)[^1S_0]_{\bf\bar 3}$ and
$(bc)[^1S_0]_{\bf 6}$ of the gluon-gluon fusion mechanism. As for
mgenxi=3, it is to generate mixed events of $(bb)[^3S_1]_{\bf\bar
3}$ and $(bb)[^1S_0]_{\bf 6}$ of the gluon-gluon fusion mechanism.
In default, we set imix=0.

\item  ixiccstate=: symbol for producing the events
from different mechanisms and different intermediate diquark states
in the case of imix=0. When it is used for the production of
$\Xi_{cc}$, the option of ixiccstate is $\in[1,6]$ : ixiccstate=$1$
stands for the case of gluon-gluon fusion with the intermediate
diquark in $(cc)[^3S_1]_{\bf\bar 3}$ state; ixiccstate=$2$ stands
for the case of gluon-gluon fusion with the intermediate diquark in
$(cc)[^1S_0]_{\bf 6}$ state; ixiccstate=$3$ stands for the case of
gluon-charm collision with the intermediate diquark in
$(cc)[^3S_1]_{\bf\bar 3}$ state; ixiccstate=$4$ stands for the case
of gluon-charm collision with the intermediate diquark in
$(cc)[^1S_0]_{\bf 6}$ state; ixiccstate=$5$ stands for the case of
charm-charm collision with the intermediate diquark in
$(cc)[^3S_1]_{\bf\bar 3}$ state; ixiccstate=$6$ stands for the case
of charm-charm collision with the intermediate diquark in
$(cc)[^1S_0]_{\bf 6}$ state. When it is used for the production of
$\Xi_{bb}$, the option of ixiccstate is $\in[1,2]$: ixiccstate=$1$
stands for the case of gluon-gluon fusion with the intermediate
diquark in $(bb)[^3S_1]_{\bf\bar 3}$ state; ixiccstate=$2$ stands
for the case of gluon-gluon fusion with the intermediate diquark in
$(bb)[^1S_0]_{\bf 6}$ state. When it is used for the production of
$\Xi_{bc}$,  the option of ixiccstate is $\in[1,4]$ : ixiccstate=$1$
stands for the case of gluon-gluon fusion with the intermediate
diquark in $(bc)[^3S_1]_{\bf\bar 3}$ state (for $\Xi_{1bc}$);
ixiccstate=$2$ stands for the case of gluon-gluon fusion with the
intermediate diquark in $(bc)[^1S_0]_{\bf 6}$ state (for
$\Xi_{1bc}$); ixiccstate=$3$ stands for the case of gluon-gluon
fusion with the intermediate diquark in $(bc)[^3S_1]_{\bf 6}$ state
(for $\Xi_{2bc}$); ixiccstate=$4$ stands for the case of gluon-gluon
fusion with the intermediate diquark in $(bc)[^1S_0]_{\bf\bar{3}}$
state (for $\Xi_{2bc}$). And its value should be set according to
one's needs.

\item  imixtype=: the type of mixed double-heavy baryon ($\Xi_{cc}$,
$\Xi_{bc}$, $\Xi_{bb}$) events for the same mechanism. It is used
when imix=1. In the case of $\Xi_{cc}$, imixtype $\in [1,3]$:
imixtype=1 is to generate the events for the gluon-gluon fusion
mechanism with the summed up results of the intermediate diquark in
$(cc)[^3S_1]_{\bf\bar 3}$ and $(cc)[^1S_0]_{\bf 6}$ states;
imixtype=2 is to generate events for the gluon-charm collision
mechanism with the results of the intermediate diquark in
$(cc)[^3S_1]_{\bf\bar 3}$ and $(cc)[^1S_0]_{\bf 6}$ states summed
up; imixtype=3 is to generate events for the charm-charm collision
mechanism with the results of the intermediate diquark in
$(cc)[^3S_1]_{\bf\bar 3}$ and $(cc)[^1S_0]_{\bf 6}$ states summed
up. In the case of $\Xi_{bc}$ or $\Xi_{bb}$, only one mixed type of
event is programmed, i.e. imixtype$\equiv$1 is to generate the
events for the gluon-gluon fusion mechanism with the results of the
intermediate diquark in $(bc)[^3S_1]_{\bf\bar 3}$, $(bc)[^1S_0]_{\bf
6}$, $(bc)[^3S_1]_{\bf 6}$ and $(bc)[^1S_0]_{\bf \bar 3}$ states for
the production of $\Xi_{bc}$ summed up or those with the
intermediate diquark $(bb)[^3S_1]_{\bf\bar 3}$ and $(bb)[^1S_0]_{\bf
6}$ states for the production of $\Xi_{bb}$ summed up. And its value
should be set according to one's needs.

\item  ivegasopen=: (status of VEGAS
subroutine) ivegasopen=$1$ means the VEGAS subroutine is `on';
ivegasopen=$0$ means the VEGAS subroutine is `off'. In default, we
set ivegasopen=1.

\item  igrade=: whether to use the grade generated by
previous VEGAS running, which is used only in the case of
ivegasopen=$0$. igrade=$1$ means to use; igrade=$0$ means not to
use. Thus one runs VEGAS once enough, since one can use the existed
grade to generate events by setting ivegasopen=$0$ (without running
VEGAS) and igrade=$1$. In default, we set igrade=0.

\item  iveggrade=: whether to improve the existed
grade generated by the previous (earlier) VEGAS running in the case
ivegasopen=$1$. igrade=$1$ means to use; igrade=$0$ means not to
use. By setting igrade=$1$, one can generate a more precise grade
based on the existed grade. Such option is used in the cases when
one does not satisfy with the MC precision of the old existed grade.
In default, we set iveggrade=0.

\item  number=: total number of times for calling the
integrand. The parameter is needed only when ivegasopen=$1$. Its
value, together with the value of the following parameter nitmx,
should be adjusted according to one's precision goal.

\item  nitmx=: upper limit for the number of
iterations. The parameter is needed only when ivegasopen=$1$.

\item  nev=: number of the events to be generated for the hadronic
production of $h+h\to \Xi_{cc}+\cdots$, $h+h\to \Xi_{bc}+\cdots$ or
$h+h\to \Xi_{bb}+\cdots$ ($h$ means a hadron) respectively according
to the value of mgenxi. Its value should be set according to one's
needs.

\item  ioutpdf=: indicates whether the three latest versions of PDFs:
CTEQ6HQ \cite{6hqcteq}, GRV98L \cite{98lgrv} and MRST2001L
\cite{2001lmrst} are used when generating the events. The three PDFs
are offered in the program for the hadronic production. ioutpdf=$1$
means they are used with further option ipdfnum, while ioutpdf=$0$
means they are not used so the inner PDFs of PYTHIA are used by the
option mstp(51). In default, we set ioutpdf=1.

\item  mstp(51)=: choice of the proton parton-distribution set
(PYTHIA parameter, see PYTHIA manual), e.g. mstp(51)=$2$ for CTEQ3M,
MSTP(51)=$7$ for CTEQ5L, mstp(51)=$8$ for CTEQ5M, and etc.. It is
used when ioutpdf=0 and in default, we set mstp(51)=7.

\item  ipdfnum=: indicates which one of the three
latest version of the PDFs CTEQ6HQ \cite{6hqcteq}, GRV98L
\cite{98lgrv} and MRST2001L \cite{2001lmrst} is used for the
hadronic production. It comes into operation only under the option
ioutpdf=$1$. ipdfnum=$100$ means to use GRV98L; ipdfnum=$200$ means
to use MRST2001L; ipdfnum=$300$ means to use CTEQ6HQ. It should be
noted that only CTEQ6HQ is consistent with the calculation
technology of GM-VFN scheme, which is our default choice for PDF.
The other two PDFs GRV98L and MRST2001L programmed here are only for
comparison with the results in literature easily. In default, we set
ipdfnum=100.

\item  idwtup=: master switch dictating how the event
weights and the cross-sections should be interpreted (PYTHIA
parameter, see PYTHIA manual). When idwtup=$1$ means, parton-level
events have a weight at the input to PYTHIA. The events generated
stochastically are either accepted or rejected according to the
weight, so that fully generated events at the output have a common
weight; parton-level events have a unit weight at the input when
idwtup=$3$, $i.e.$, they are always accepted \footnote{For the
present usages, of PYTHIA the two options idwtup=$1$ and idwtup=$3$
are enough.}. In default, we set idwtup=3.

\item  mstu(111)=: order of $\alpha_s$ in the
evaluation in PYALPS (a PYTHIA routine for calculating $\alpha_s$,
see PYTHIA manual); $e.g.$, mstu(111)=$1$ for leading order (LO);
mstu(111)=$2$ for next leading order (NLO). In default, we set
mstu(111)=1.

\item  paru(111)=: constant value of $\alpha_s$ (see
PYTHIA manual), which is used only when mstu(111)=$0$. In default,
we set paru(111)=0.2.

\item isubonly=: whether to generate the information only of the
gluon-gluon fusion subprocess; isubonly=$0$ for the full hadronic
production, i.e. the structure functions are connected; isubonly=$1$
for the subprocess only. In default, we set isubonly=0.

\item subenergy=: the energy (in units GeV) of the gluon-gluon fusion
subprocess. It is needed only when isubonly=$1$. And its value
should be set according to one's needs.

\end{itemize}

\subsection{Check of the generator}

\begin{figure} \centering
\includegraphics[width=0.45\textwidth]{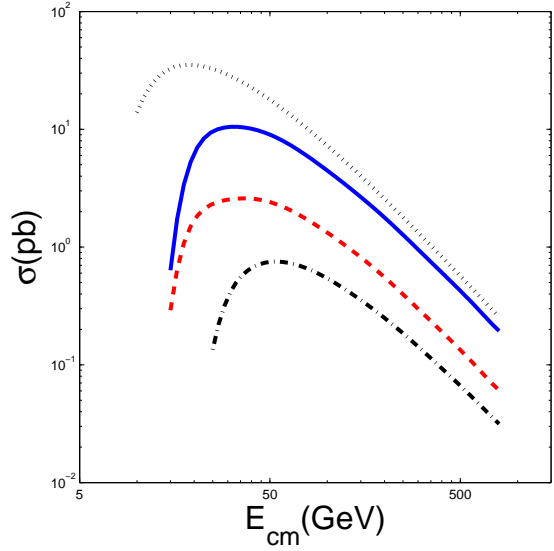}
\caption{The energy dependence of the integrated partonic
cross-section for the production of the double-heavy baryons with
heavy diquarks via the gluon-gluon fusion mechanism. The dotted
line, solid line, dashed line and dash-dot line stand for the
baryons with $(cc)_{\bf\bar{3}}[^3S_1]$, $(bc)_{\bf\bar{3}}[^3S_1]$,
$(bc)_{\bf\bar{3}}[^1S_0]$ and $(bb)_{\bf\bar{3}}[^3S_1]$
respectively. Note here that the curves for $\Xi_{cc}$ and
$\Xi_{bb}$ both are divided by a factor `2' for convenience in
comparison with literature results.} \label{subcs}
\end{figure}

First of all, all the programs are checked by examining the gauge
invariance of the amplitude, i.e. the amplitude vanishes when the
polarization vector of an initial/final gluon is substituted by the
momentum vector of this gluon. Numerically, we find that the gauge
invariance is guaranteed at the computer ability (double precision)
for all these processes. Especially, to make sure the correctness of
our program for the most complicated gluon-gluon fusion mechanism,
we have checked whether the numerical results evaluated by the
generator agree well with those by FDC package \cite{fdc,cqww}. For
such purpose, we have added the parameter {\bf isubonly} to decide
whether to generate the information only of the gluon-gluon fusion
subprocess, since it is much more easier to be compared.
Furthermore, we have compared the numerical results for the
gluon-gluon fusion mechanism with those by using the same input
parameters as stated in the corresponding references in the
literature, and the examples of these comparison can be explicitly
found in the section III of Ref.\cite{cqww}. In checking the
program, we have recalculated all the values listed in
Ref.\cite{cqww}, and we find a good agreement of our present results
with those of Ref.\cite{cqww}. More explicitly, we obtain the same
results for the partonic cross sections as that of Ref.\cite{cqww}
and we show the partonic cross sections for the production of
baryons with heavy diquarks via the gluon-gluon fusion subprocess in
Fig.(\ref{subcs}).

\begin{figure} \centering
\includegraphics[width=0.5\textwidth]{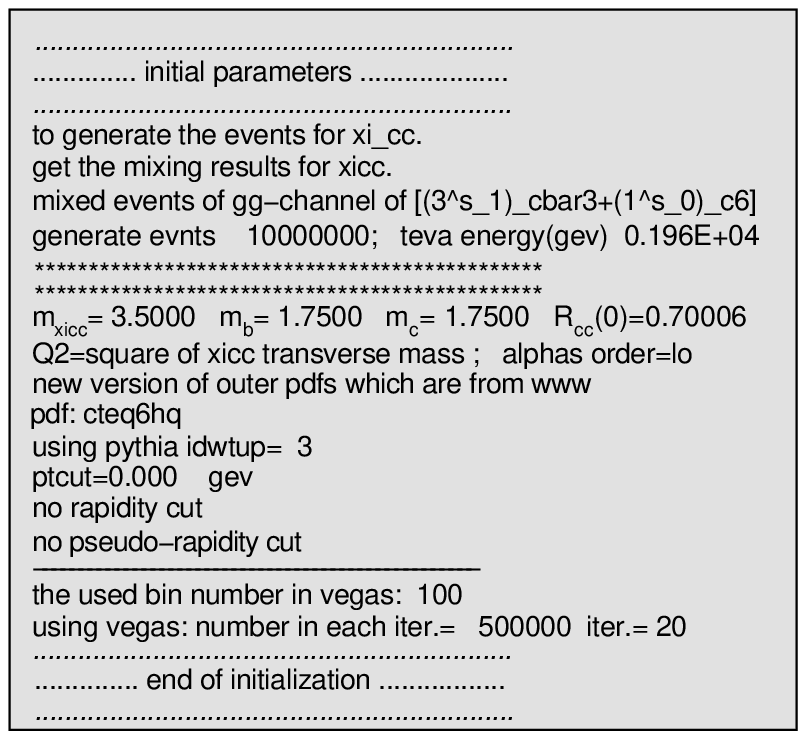}
\caption{Snapshot of the initial parameters used in the test run of
the generator GENXICC, which is for the hadronic production of
$\Xi_{cc}$ at the TEVATRON and is to generate the mixed events
through gluon-gluon fusion mechanism (imix=1 and imixtype=1).}
\label{test}
\end{figure}

\begin{figure}
\centering
\includegraphics[width=0.48\textwidth]{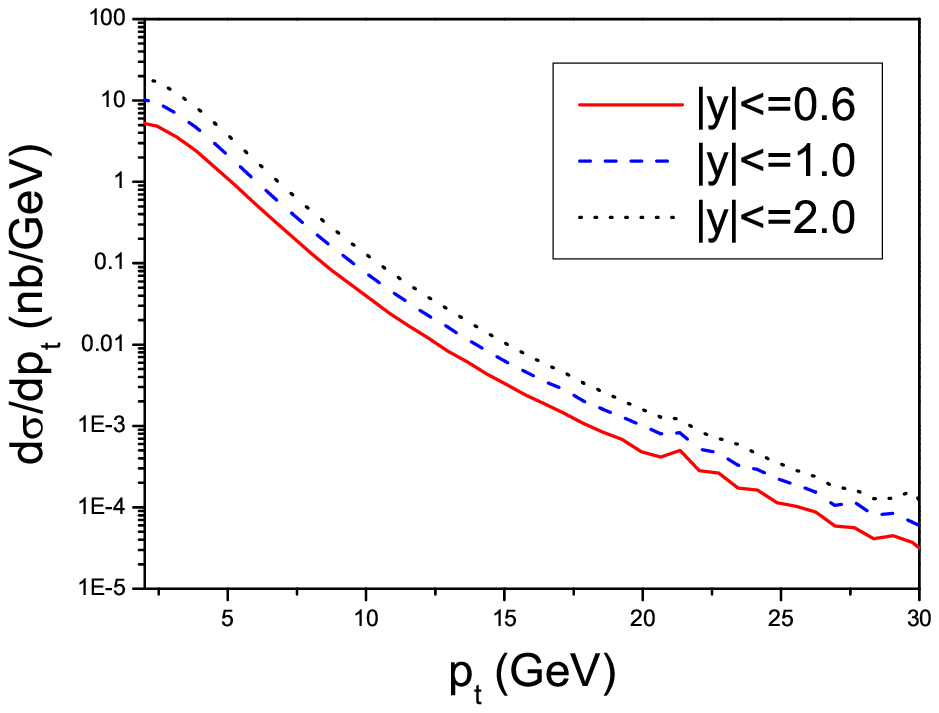}
\includegraphics[width=0.49\textwidth]{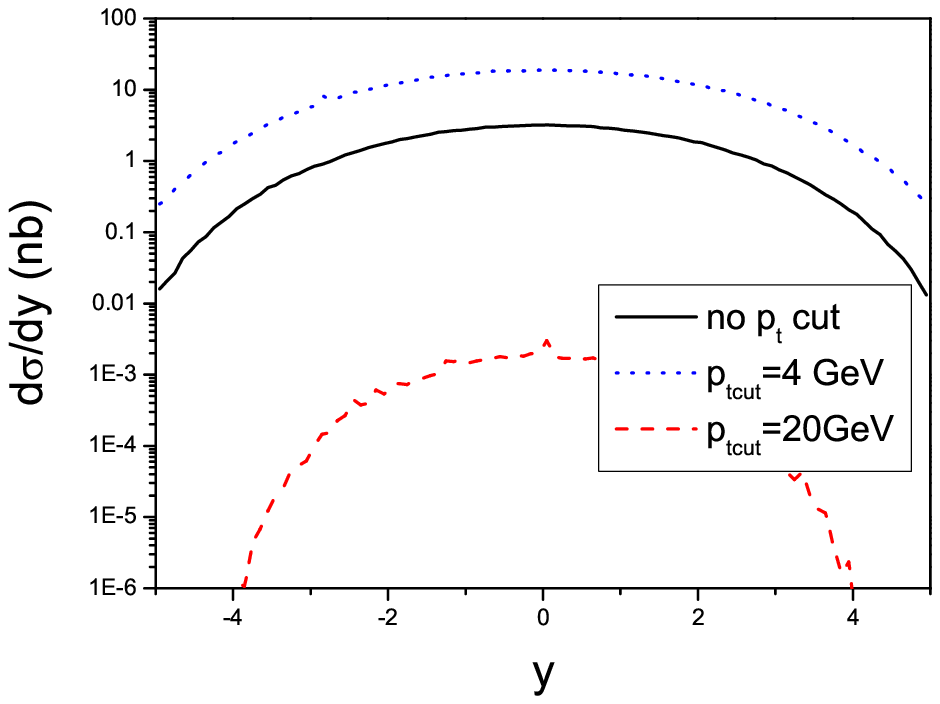}
\caption{$\Xi_{cc}$-$p_t$ and $\Xi_{cc}$-$y$ distributions for the
test run of the generator GENXICC, which is for the hadronic
production of $\Xi_{cc}$ at the TEVATRON and is to generate the
mixed events through gluon-gluon fusion mechanism (imix=1 and
imixtype=1).} \label{distribution}
\end{figure}

Finally, we show a test run for the hadronic production of
$\Xi_{cc}$ at the TEVATRON. When running the program, the initial
parameters are shown in Fig.(\ref{test}). All the obtained data are
turned into a zipped file, named as testdata.tar.gz in the main
directory of the program. Some typical resultant curves, e.g. $p_t$-
and $y$-distributions are shown in Fig.(\ref{distribution}). In
drawing the distributions, several typical rapidity cut, e.g.
$|y_{cut}|=0.6$, $0.1$ and $0.2$ are taken for $\Xi_{cc}$-$p_t$
distributions and similarly, several typical $p_{tcut}$ are taken
for $\Xi_{cc}$-$y$ distributions.

\section{Summary}

A generator GENXICC has been developed and well-tested by us, which
is for hadronic production of $\Xi_{cc}$ based on three mechanisms
via gluon-gluon fusion, gluon-charm collision and charm-charm
collision. While for the hadronic production of $\Xi_{bc}$ and
$\Xi_{bb}$ only the most important mechanism via gluon-gluon fusion
is available, because we think that considering the characteristics
and most potential usages it is enough, and the program may be
simplified in certain degree. The generator has proper interface to
PYTHIA, so the full events of the production can be generated by
filling the standard PYTHIA event common block. Note that in the
program we do not distinguish the state $\Xi_{QQ'}$ (in spin one
half) and the state $\Xi^*_{QQ'}$ (in spin three halves) and
consider their production of them together. If one would like to
know their difference of the production of two states, roughly
speaking, one can estimate it by the way to account the production
as that if it is via $^1S_0$ heavy diquark then all (100\%) of the
diquarks will be fragmented into the state $\Xi_{QQ'}$, while if it
is via $^3S_1$ heavy diquark then $\frac{1}{3}$ fraction of the
diquarks will be fragmented into the state $\Xi_{QQ'}$ and
$\frac{2}{3}$ fraction will be fragmented into the state
$\Xi^*_{QQ'}$. In view of the prospects for double-heavy baryon
($\Xi_{cc}$, $\Xi_{bc}$ and $\Xi_{bb}$) physics at SELEX, Tevatron
and LHC, the generator offers a useful tool for further experimental
feasibility studies of the double-heavy baryons.

\begin{center}

\vspace{2cm}

{\bf ACKNOWLEDGEMENTS}
\end{center}
This work was supported in part by the Natural Science Foundation of
China. X.G. Wu would like to thank the support from the China
Postdoctoral Science Foundation. \\


\begin{thebibliography}{s2}

\bibitem{exp} M. Mattson {\it et al.}, SELEX Collaboration,
Phys. Rev. Lett. {\bf 89}, 112001(2002).

\bibitem{exp2} A. Ocherashvili {\it et al.}, SELEX Collaboration,
hep-ex/0406033.

\bibitem{baranov} S.P. Baranov, Phys. Rev. D{\bf 54}, 3228(1996).

\bibitem{kiselev1} A.V. Berezhnoy, V.V. Kiselev, A.K. Likhoded
and A.I. Onishchenko, Phys. Rev. D{\bf 57}, 4385(1998).

\bibitem{cqww} Chao-Hsi Chang, Cong-Feng Qiao, Jian-Xiong Wang and Xing-Gang
Wu, Phys. Rev. D{\bf 73}, 094022(2006).

\bibitem{cmqw} Chao-Hsi Chang, Jian-Ping Ma, Cong-Feng Qiao and Xing-Gang
Wu, hep-ph/0610205.

\bibitem{pythia} T. Sjostrand, Comput. Phys. Commun. {\bf 82}, 74
(1994); T. Sjostrand, S. Mrenna and P. Skands, JHEP {\bf 0605},
026(2006); T. Sjostrand, S. Mrenna and P. Skands, hep-ph/0603175.

\bibitem{brodsky} S.J. Brodsky, P. Hoyer, C. Peterson and N. Sakai,
Phys. Lett. B {\bf 93}, 451 (1980); S. J. Brodsky, C. Peterson and
N. Sakai, Phys. Rev. D{\bf 23}, 2745 (1981); R. Vogt and S.J.
Brodsky, Nucl. Phys. B{\bf 478}, 311(1996).

\bibitem{bcvegpy1} Chao-Hsi Chang, Chafik Driouich, Paula Eerola and Xing-Gang Wu,
Comput. Phys. Commun. {\bf 159}, 192 (2004); hep-ph/0309120.

\bibitem{bcvegpy2} Chao-Hsi Chang, Jian-Xiong Wang and Xing-Gang
Wu, Comput. Phys. Commun. {\bf 174}, 241 (2006); hep-ph/0504017.

\bibitem{bcvegpy3} Chao-Hsi Chang, Jian-Xiong Wang and Xing-Gang
Wu, Comput. Phys. Commun. {\bf 175}, 624 (2006); hep-ph/0604238.

\bibitem{nrqcd} G.T. Bodwin, E. Braaten, and G.P. Lepage,
Phys. Rev. D51, 1125 (1995); Erratum:{\it ibid.}, D55, 5853 (1997).

\bibitem{gmvfn1} M.A.G. Aivazis, J.C. Collins, F.I. Olness and W.K. Tung,
Phys. Rev. D{\bf 50}, 3102(1994); Phys. Rev. D{\bf 50}, 3085(1994);
F.I. Olness, R.J. Scalise and W.T. Tung, Phy. Rev. D{\bf 59},
014506(1998).

\bibitem{gmvfn2} J. Amundson, C. Schmidt, W.K. Tung and X.N. Wang,
JHEP10, 031(2000).

\bibitem{6hqcteq} S. Kretzer, H.L. Lai, F.I. Olness and W.K. Tung,
Phys. Rev. D{\bf 69}, 114005(2004).

\bibitem{98lgrv} M. Glueck, E. Reya, A. Vogt, Eur. Phys. J. C {\bf 5},
461 (1998).

\bibitem{2001lmrst} A.D. Martin, R.G. Roberts, W.J. Stirling and R.S. Thorne,
Eur. Phys. J. C {\bf 23}, 73 (2002).

\bibitem{vegas} G.P. Lepage, J. Comp. Phys {\bf 27}, 192 (1978).

\bibitem{rambos} R. Kleiss and W.J. Stirling, Comput. Phys. Commun. {\bf
40}, 359(1986).

\bibitem{formcalc} T. Hahn and M. Rauch, hep-ph/0601248(2006).

\bibitem{majp} J.P. Ma and Z.G. Si, Phys. Lett. B{\bf 568},
135(2003).

\bibitem{ccfl} Chao-Hsi Chang, Shao-Long Chen, Tai-Fu Feng and
Xue-Qian Li, Phys.Rev. D{\bf 64}, 014003(2001).

\bibitem{fdc} Jian-Xiong Wang, Nucl. Instrum. Methods Phys. Res.,
Sect. A 534, 241 (2004).

\end{thebibliography}
\end{document}